\documentclass[10pt,conference]{IEEEtran}
\makeatletter
\IEEEoverridecommandlockouts

\def\ps@headings{%
\def\@oddhead{\mbox{}\scriptsize\rightmark \hfil \thepage}%
\def\@evenhead{\scriptsize\thepage \hfil \leftmark\mbox{}}%
\def\@oddfoot{}%
\def\@evenfoot{}}
\makeatother 

\usepackage{color}
\usepackage{graphicx}               
\usepackage{algorithm}
\usepackage{algpseudocode}
\usepackage[cmex10]{amsmath}
\usepackage{soul}

\ifCLASSINFOpdf
\else
\fi

\usepackage{algorithm}

\usepackage{algpseudocode}
\usepackage{caption}

\usepackage{url}

\usepackage{algorithm}
\usepackage{algpseudocode}

\begin{document}
%
\title{Decentralized Load Management in HAN: \\An IoT-Assisted Approach \vspace{-0.3cm}}

\author{\IEEEauthorblockN{Jagnyashini Debadarshini, Sudipta Saha, Subhransu Ranjan Samantaray}
\IEEEauthorblockA{School of Electrical Sciences,
Indian Institute of Technology Bhubaneswar}
\IEEEauthorblockA{Email: \{\emph{jd12, sudipta, srs}\}@iitbbs.ac.in}
}

\maketitle



%
\IEEEpeerreviewmaketitle

\begin{abstract}

A \textit{Home Area Network} (HAN) is considered to be a significant component of \textit{Advanced Metering Infrastructure} (AMI) and has been studied well in many works. It binds all the electrical components installed in a defined premise together for their close monitoring and management. However, HAN has been realized so far mostly as a centralized system. Therefore, like any other centralized system, the traditional realization of HAN also suffers from various well-known problems, such as single-point-of-failure, susceptibility to attacks, requirement of specialized infrastructure, inflexibility to easy expansion, etc. To address these issues, in this work, we propose a decentralized design of HAN. In particular, we propose an IoT based design where instead of a central controller, the overall system operation is controlled and managed through decentralized coordination among the the electrical appliances. We leverage \textit{Synchronous-Transmission} (ST) based data-sharing protocols in IoT to accomplish our goal. To demonstrate the efficacy of the proposed decentralized framework, we also design a real-time intra-HAN load-management strategy and implement it in real IoT-devices. Evaluation of the same over emulation platforms and IoT testbeds show upto 62\% reduction of peak load over a wide variety of load profiles.
\end{abstract}


\section{Introduction}
\label{sec:intro}
\begin{figure*}[htbp]
\begin{center}
\includegraphics[width=\linewidth]{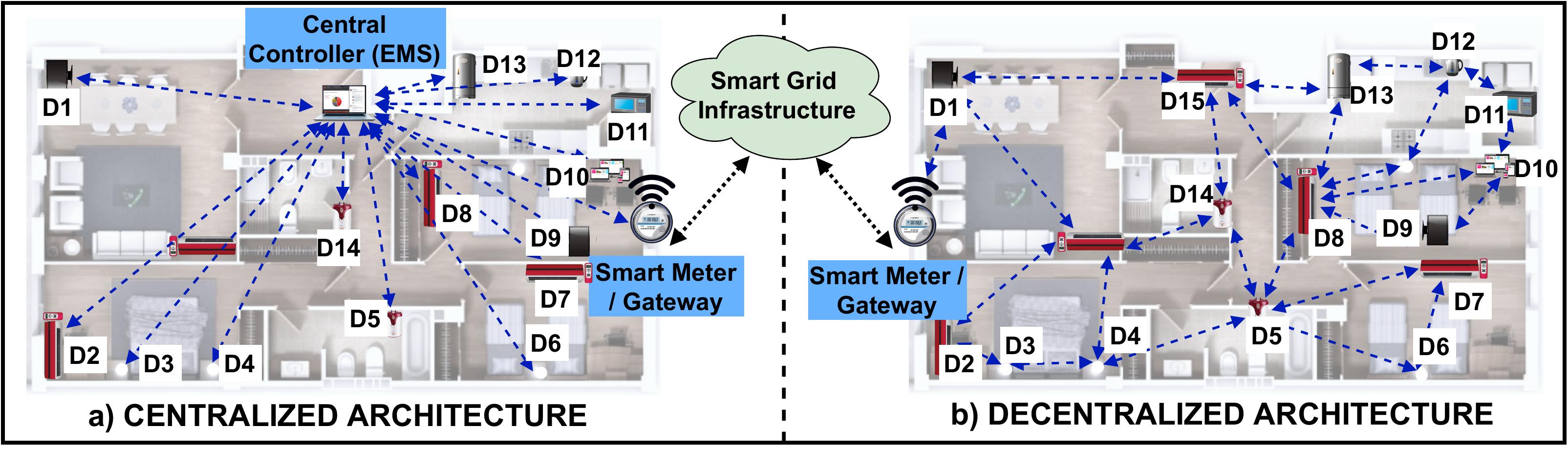}
\end{center}
\vspace{-0.3cm}
\caption{(a) A centralized HAN where the electrical appliances D1, D2, ..., D14 are directly connected to the central-controller (\textit{EMS}). (b) A decentralized design of the same HAN where all the devices are connected with each other through low-power links forming a multi-hop connected network where Smart-Meter (Gateway).}
\label{fig:cendecent}
\end{figure*}
Monitoring of the consumption of electricity and its management are among the primary issues in an \textit{Electric Power Distribution System}. In a smart-grid, the sub-system known as the \textit{Advanced Metering Infrastructure} (AMI) \cite{ghosal2019key} is primarily responsible for this crucial task. A \textit{Home Area Network} (HAN) \cite{yan2012survey} is regarded as the end-point of an AMI. A HAN binds together all the electrical equipment installed within a certain defined premise under a single \textit{local area network} for tight monitoring and management of their operations \cite{namboodiri2013toward}. The performance of an AMI, hence largely depends on the collective performance of all the HANs.

Most of the current implementations of HAN follow a centralized architecture \cite{caprino2014peak,lee2011power,6476768,kim2011scheduling,mohsenian2010optimal} where all the electrical appliances are connected to a central controller (see Figure \ref{fig:cendecent}(a)). However, a centralized solution, in general, bears a number of drawbacks. The root cause of all these problems is mainly the possibility of \textit{single-point-of-failure} in any centralized system. In a HAN, failure in the central controller would result in a complete halt of the whole system. Being a very crucial unit of smart-grid, a HAN-controller is, thus, also a very natural target of the adversaries. A compromised HAN controller would literally let the whole premise to be remotely controlled by an adversary. Moreover, a set of compromised HANs can easily create a strong impact even over the whole smart-grid system.

In addition, centralized systems are in general \textit{not flexible} and also not \textit{easily expandable} due to certain capacity limit of the deployed central-controller. Traditional centralized HAN solutions would also involve costly installation inside the premise, e.g., costly central-controller as well as installation of a suitable wired or wireless LAN infrastructure, etc. Moreover, any wireless communication activity in an indoor environment always gives rise to the radiation level and hence any such solution, specially those which are supposed to be constantly used throughout a day, should attempt to reduce the radiation level as much as possible. Under a centralized setting this is definitely harder to achieve.

In this work we propose a low-power IoT-based decentralized model of HAN which can effectively solve many of the above mentioned problems. However, note that, in a centralized model, the central-controller takes most of the crucial decisions. On the contrary, in a decentralized model, the participating entities themselves derive the decisions through collaboration and cooperation. Therefore, to effectively replace a centralized model with a decentralized one what is necessary is an efficient strategy that can enable easy, fast and reliable coordination among the electrical appliances which can in turn be leveraged to satisfy all the other needs. In this work we exploit the recent advancements in the IoT-technology and the low-power communication protocols to build the proposed decentralized model of HAN.

There have been immense development in the IoT-technology, especially in the communication protocols. In many recent works \cite{zimmerling2020synchronous} \textit{Synchronous-Transmission} (ST) based communication protocols have been proved to be performing much better than the traditional \textit{Asynchronous-Transmission} (AT) based strategies. In general, a ST based protocol, try to avoid the collision among transmission of the packets from different source nodes as much as possible through clever exploitation of special physical layer phenomena known as \textit{Capture-Effect} (CE)/ \textit{Constructive-Interference} (CI). This is achieved through a system wide time-synchronization among the transmitters. Because of substantial reduction in the loss of time and energy through avoidance of collision, ST based protocols successfully achieve higher reliability under low latency. In this work, we propose an ST-based framework as a foundation for the proposed decentralized HAN. We leverage the same to develop an efficient and fast intra-HAN scheduling strategy to manage the peak load and flatten the load curve through efficient collaboration among the electrical appliances.

The contributions from the work are summarized below.
\begin{itemize}
    \item We propose a lightweight, cost-effective, flexible design of a decentralized HAN where the electrical appliances can seamlessly collaborate with each other and accomplish their goals without any intervention from any central controller.
    \item We leverage \textit{Synchronous-Transmission} (ST) based IoT-protocols for design and development of a lightweight and decentralized strategy for scheduling of the electrical appliances for efficient load-management.
    \item We implement our proposed design in Contiki Operating System for TelosB devices. Through extensive experiments in emulation and IoT-testbeds we demonstrate the effectiveness of the proposed design.
\end{itemize}

The rest of the paper is organized as follows. In Section \ref{sec:related} and \ref{sec:back} we present the related works and a brief background on ST, respectively. The details of the design of the proposed decentralized HAN and the scheduling strategy are presented in Section \ref{sec:design}. Section \ref{sec:simulation} presents the evaluation study of the proposed strategy in emulation platform and testbed. Finally Section \ref{sec:conclusion} concludes the work with some future directions.

\section{Related Works}
\label{sec:related}

Centralized architecture of HAN has been quite commonly used in many existing works. All these works specifically talk about a special central controller which is in charge of managing and controlling all the operations in a HAN. For example \textit{Centralized Control Station} \cite{caprino2014peak,lee2011power,6476768,kim2011scheduling,mohsenian2010optimal}, \textit{Building Energy Management System} \cite{missaoui2014managing}, \textit{Dynamic Demand Response Controller} \cite{yoon2014dynamic}, \textit{Energy Box} \cite{agnetis2013load}, \textit{Smart Scheduler} \cite{adika2013autonomous} etc.

A centralized architecture naturally brings a very good grip over the full system which immensely help in realizing various advanced scheduling strategies. For instance, the work \cite{caprino2014peak} uses earliest deadline first scheduling algorithm for reducing the peak load. The work \cite{lee2011power} specially considers the dynamics of the preemptive and non-preemptive tasks in their design.
The works \cite{6476768,kim2011scheduling} exploit 
stochastic scheduling strategies. Real-time pricing as well as waiting time of the devices have been considered in the design proposed in the work \cite{mohsenian2010optimal}. The work \cite{agnetis2013load} considers the climatic factors along with cost minimization and exploits heuristic optimization based methods for scheduling.

However, a centralized architecture inherently suffers from a series of issues as already described in Section \ref{sec:intro}. Therefore, in this work we propose a decentralized design of a HAN as an alternative. In addition, we explore how a decentralized design can also foster ways to support strong control and management capability over the electrical appliances connected by a HAN through collaboration and coordination. In the next section we provide description of the ST based protocols that are used as primary tools to accomplish our goals.

\section{Background}
\label{sec:back}

ST have been used in many recent works to serve various purposes related to information-management in IoT and WSN systems, e.g., data-sharing, data-collection, data-aggregation \cite{zimmerling2020synchronous}. It has been shown that the performance of the ST based protocols are much better than the AT based ones. Unlike AT, the protocols under ST in general make the nodes more cooperate than compete with each other.
To accomplish the goals ST based protocols exploits network wide time-synchronization. The pioneering work Glossy \cite{glossy} demonstrates a lightweight strategy to achieve the same as well as show that ST can be exploited to achieve quick network wide one-to-all flooding of data. However to achieve our goal, what is fundamentally necessary is a mechanism through which all the electrical appliances connected under a HAN can reliably share their status/control data with each other in real-time and thereafter apply appropriate logic to derive crucial decisions as necessary without help from any central controller. To achieve this, we use a ST based all-to-all data-sharing strategy, \textit{MiniCast} as described briefly below.

\textbf{MiniCast}: MiniCast \cite{saha2017efficient} allows multiple source nodes to quickly share their data with each other. In particular, it efficiently extends the functionality of Glossy. The protocol starts with a designated node called \textit{initiator} transmitting the first packet (probe) that triggers the transmission of the packets from the neighbors of the initiator as per a predefined TDMA schedule. The transmissions are done in the form of a chain of packets marked by two special packets HDR and TRL for staring and ending, respectively. The transmission of the chain from the first-hop nodes triggers the transmission from the second hop-nodes and so on. Note that all the transmissions are time-aligned in such a way that the protocol can get benefited by CE/CI \cite{leentvaar1976capture} and the packets get properly received in the corresponding destinations. Fundamentally there are several other solutions for many-to-many/all-to-all data-sharing in IoT/WSN \cite{zimmerling2020synchronous}. However, unlike others, MiniCast exploits a special form of CE where the packets that are transmitted in a time-aligned manner bear exactly the same content. This is referred to as Homogeneous CE which makes MiniCast perform much better than other solutions even under very dense network setting. A HAN network is usually considered to be densely packed with many electrical appliances within a confined premise. Therefore, in this work we select MiniCast as our primary tool.
\section{Design}
\label{sec:design}
In the following we first provide brief description of various components of the proposed design followed by the details of the load-management protocol.

\textbf{Device-Interface:} Every electrical-appliance that is to be controlled and managed through the HAN is required to have two interfaces. \textit{First}, an usual \textbf{Electrical-Interface} (EI) which connects the appliance with the electricity supply. \textit{Second}, a \textbf{Device-Interface} (DI), which enables the connectivity of the appliance with the HAN. A DI is designed based on a standard low-power IoT/WSN device. It is supposed to have a set of input and output lines to control the associated electrical appliance. The input lines are used to read the status of the connected appliance while the output lines are used to send the necessary instruction (e.g., ON, OFF) to the appliance.

\textbf{Types of Appliances}: We divide the electric appliances into two categories - \textit{Type-1} and \textit{Type-2}. Type-1 appliances consume comparatively lesser power than Type-2. For example, light-bulb, electric-fan, charger, TV, etc fall under Type-1 while fridge, air-condition, room-heater, microwave-oven, water-heater, washing-machine, cloth-dryer etc., fall under Type-2. 
Moreover, the users of Type-1 appliances usually need a very prompt service while the same may not be always true for the Type-2 appliances. In fact exact timing of the execution of the Type-2 appliances can be deferred little bit without hampering user satisfaction. Thus, Type-2 appliances have the flexibility of getting scheduled in a desired way which is not possible with Type-1 appliances. We exploit this facility in our load management protocol.

\textbf{Duty-Cycle:} Almost all the Type-2 appliances are designed in such a way that they can \textit{duty-cycle} their main electrical component that consumes the maximum amount of power, e.g., the compressor in AC/fridge. In particular, these appliances keep the main electrical unit ON for a certain duration and then turn it OFF. This is repeated periodically. How long to stay ON and OFF are decided based on the desired goal, e.g, the exact room temperature to be set by an AC. It also depends on external factors such as environmental temperature etc. Although modern ACs are designed in a way so that the overall operations happen in a very smooth way, but internally the duty-cycle concept is still followed. 

\textbf{maxDCP and minDCD}: Duty-cycling has mainly two components - \textit{duty-cycle period} and \textit{duty-cycle duration}. The former one is the period of the cyclic process and the later one is the duration for which the main unit of an appliance is kept ON in each of the cycles. The lesser the duty-cycle duration (for a given duty-cycle period) the lesser the consumption of electricity by the appliance. However, to satisfy a specific requirement, an appliance must run for minimum duty-cycle duration which we refer to as \textit{minDCD}. Similarly, the more duty-cycle period (for a give duty-cycle duration), the lesser the consumption of power. However, in practice to serve a certain purpose, the duty-cycle period must not go beyond a certain max value which we refer to as \textit{maxDCP}.
\begin{figure}[htbp]
\begin{center}
\includegraphics[angle=0,width=0.48\textwidth]{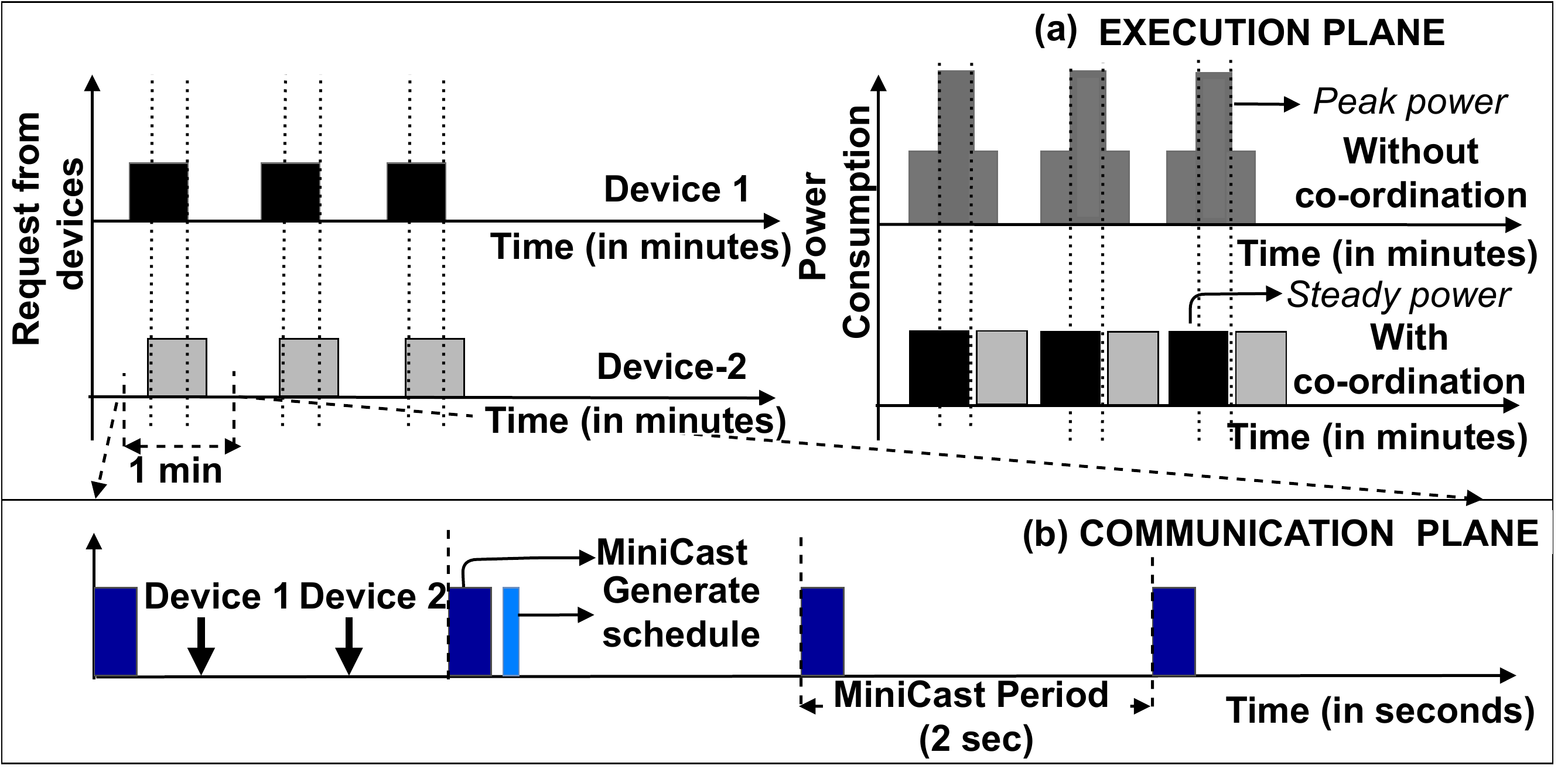}
\vspace{-0.2cm}
\end{center}
\caption{(a) Execution-Plane (EP) where two devices execute their operation with coordination (bottom) and without-any coordination (above). (b) Communication-Plane (CP) where DIs communicate with each other using MiniCast.}
\label{fig:concept_picture}
\vspace{-0.2cm}
\end{figure}
\\
\textbf{Load-Management}: Under an usual independent operation of the electrical appliances, the devices are free to do their own duty-cycling which naturally causes arbitrary overlap of their execution and results in sudden spike in the electrical load. In this work we try to coordinate the operation among the Type-2 appliances in such a way that the ON-OFF timing of different appliances are mutually adjusted through dynamic scheduling of their operations. 
This results in avoidance of any sudden spike or abrupt rise in the load. Rather the load rises in a step by step manner. Figure \ref{fig:concept_picture} shows this basic concept through a diagram. It demonstrates the operations in two planes: (a) \textit{Execution-Plane} (EP), where the electrical appliances carry out their \textit{jobs} (time-scale is in the order of \textit{minute}). (b) \textit{Communication-Plane} (CP), where the DIs talk to each other (time-scale is in the order of \textit{second}). Since different category of Type-2 appliances have different minDCD and maxDCP, we maintain different streams for different combination. The final electric load is the sum of all these streams together.

\textbf{Algorithm:} Algorithm \ref{algo:scheudler} describes the scheduling strategy. Users request the service by a Type-2/Type-1 appliance by pressing a button in the corresponding DI. In CP, the DIs keep on running MiniCast every second and thereby every DI comes to know the full set of requests that are not yet addressed. The algorithm uses this set of requests as the input and produces a schedule as output  where the ON/OFF status of every device is specified. It is organized in three parts, P1, P2 and P3. In P1, requests from the users are collected. P2 keeps track of the existing jobs that are not yet finished. 
P3 does the scheduling task considering all the running and waiting jobs. 
Every job is assigned a maximum time within which it must be scheduled. However, the exact schedule of the jobs depends on the number of jobs to be scheduled and the minDCD of the appliances. To minimize the peak-load we first divide the available time into a number of slots each of length minDCD. Finally, the jobs are scheduled in a round robin manner. Once the full maxDCP time is exhausted, the rest of the unscheduled appliances whose deadline is reached, are scheduled with the highest priority for in parallel execution. It increases the electric load, but in a systematic step by step manner, and hence the peak and the overall load remains under control. 


\begin{algorithm}[htbp]
\small
\caption{{ \textsc{Load-Management}}}\label{algo:scheudler}
\begin{algorithmic}
\State {\textbf{Input :} $Request[N]$}
\State {\textbf{Output :} $Status[N]$}
\end{algorithmic}
\textbf{\underline{\textsc{P1: (Request Collection)}}}\vspace{1mm}
\begin{algorithmic}[1]
\For {every device $D_{i}$}
 \If{$Request[D_i] == ON $}
    \State{$Status[D_i]=WAIT $}
    \State{$ADD(WL,D_i)$}
    \State{$wait\_time[D_i] = current\_time + MAXDCP$}
 \ElsIf{$Request[D_i] == OFF$}
    \If{$Status[D_i] == ON $}
        \State{$REMOVE(RL,D_i)$}
    \ElsIf{$Status[D_i] == WAIT $}
        \State{$REMOVE(WL,D_i)$}
    \EndIf
  \State{$Status[D_i]=OFF $}
 \EndIf
\EndFor
\vspace{2mm}
\end{algorithmic}
\textbf{\underline{\textsc{P2: (Tracking Unfinished Jobs)}}}\vspace{1mm}
\begin{algorithmic}[1]
\If {$!IS\_EMPTY(RL)$}
    \For {every device $D_{i}$ in $RL$}
        \If {$current\_time-start\_time[D_{i}] > MINDCD $ }
            \State{$Status[D_i]=WAIT $}
            \State{$REMOVE(RL,D_i)$}
            \State{$ADD(WL,D_i)$}
            \State{$wait\_time[D_i]=current\_time+MAXDCP$}
        \EndIf
    \EndFor
\EndIf
\vspace{2mm}
\end{algorithmic}
\textbf{\underline{\textsc{P3: (Scheduling)}}}\vspace{1mm}
\begin{algorithmic}[1]
\If {$!isEmpty(WL)$}
    \State{$size=FIND\_SIZE(WL)$}
    \State{$max\_wtime=FIND\_MAX\_WTIME(WL)$}
    \State{$slots=(max\_wtime-current\_time)/MINDCD$} 
    \State{$jobs\_to\_schedule =size/slots$} 
    \For {every device $D_{i}$ in $WL$}
        \If {($jobs\_to\_schedule\geq0\lor wait\_time[D_{i}]==current\_time)$}
            \State{$REMOVE(WL,D_i)$}
            \State{$ADD(RL,D_i)$}
            \State{$Status[D_i]=ON $}
            \State{$start\_time[D_{i}]=current\_time$}
            \State{$jobs\_to\_schedule=jobs\_to\_schedule-1$}
        \EndIf
    \EndFor
\EndIf
\end{algorithmic}
\end{algorithm}

\textbf{Initiator selection}: Note the role of the initiator in the proposed strategy is vastly different from that of a central-controller. Initiator is not a special node, rather, any node in the system can act as an initiator. Moreover there are algorithms for quick election of new initiator as necessary \cite{al2017network}. Due to lack of space we do not deal with these issues in details here.

\section{Evaluation}
\label{sec:simulation}
\begin{table*}
\begin{center}
\includegraphics[angle=0,width=\textwidth,height=23cm,keepaspectratio]{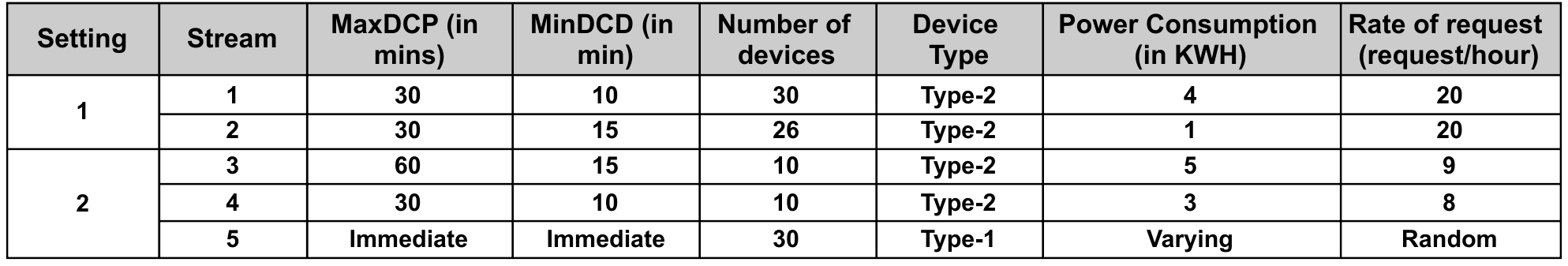}
\end{center}
\vspace{-0.3cm}
\caption{Details of the experimental settings, streams of requests and profiles of the electrical appliances.}
\vspace{-0.3cm}
\label{fig:profile}
\end{table*}
We implement the proposed load-management strategy in Contiki Operating System for TelosB devices and extensively test its operation in the Contiki network emulation system \textit{Cooja} as well as IoT/WSN testbeds. Performance of the proposed coordinated duty-cycling strategy is compared with the baseline where no coordination is used for realizing the duty-cycles in the Type-2 devices. 

\textbf{Metrics:} The comparison is done based on three metrics - (a) \textit{System-Load:} It demonstrates the overall load in the system over time. (b) \textit{Peak-Load:} It is the maximum load over a considered period. (c) \textit{Average-Load} and \textit{Standard-Deviation}: These two are used to show how the load over a given period varies with time. (d) \textit{Time Delay:} The use of the proposed strategy may keep some of the appliances in waiting state based on the exact time of their arrival. This metric measures the amount of time an appliance waits on average before it gets its first chance of execution.

\begin{figure}[htbp]
\begin{center}
\includegraphics[angle=0,width=0.48\textwidth]{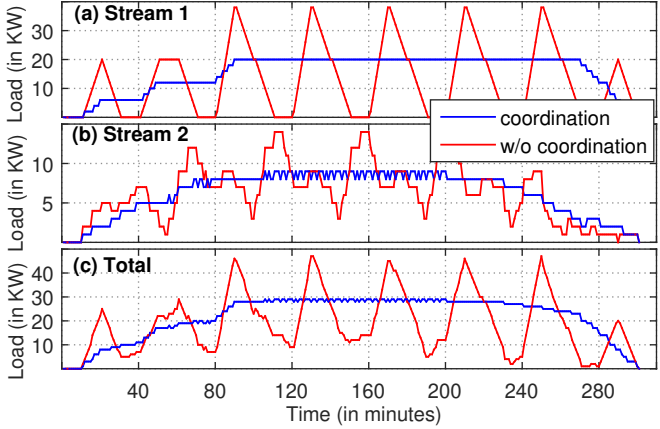}
\end{center}
\vspace{-0.3cm}
\caption{Variation of system-load for stream-1 (a), stream-2 (b), and the total load (c) for a duration of five hours. The streams are detailed in Table \ref{fig:profile}.}
\vspace{-0.6cm}
\label{fig:diff_parameter}
\end{figure}



The proposed ST based HAN is emulated in Cooja using a random network spread over a premise of size 200X200 sq. meter comprised of a number of Type-1 and Type-2 devices. Different Type-2 devices have different characteristics. We experiment with a combination of few types of such devices. For cleaner analysis we separate the requests from different category of devices in different streams. Experiments are done over two sets of streams as detailed in Table \ref{fig:profile}.

\textbf{Experiment under setting-1}: 
Setting-1 corresponds to a busy industry environment having 30 ACs (stream-1) and 26 Water-Heaters (stream-2). Requests in both the streams are assumed to be arriving at a faster rate, referred as stream-1 and stream-2 in Table-1. 
Figure \ref{fig:diff_parameter}(a) and \ref{fig:diff_parameter}(b) show the system-load in (KW) over a duration of 5 hr separately based on stream-1 and stream-2, respectively. The variation of the system-load over time is demonstrated in the results for both with and without the proposed load-management algorithm. Figure \ref{fig:diff_parameter}(c) shows the total load, i.e., when stream-1 and stream-2 are present together. In summary the proposed strategy successfully reduces the peak load upto 38\% and the overall variation in the load upto 28\% throughout the experiment.

\begin{figure}[htbp]
\begin{center}
\includegraphics[angle=0,width=0.48\textwidth]{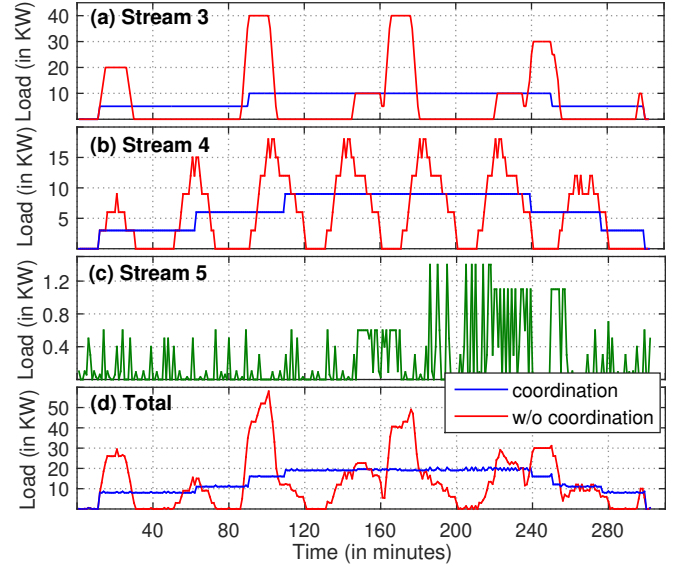}
\end{center}
\vspace{-0.2cm}
\caption{Variation of system-load for stream-3 (a), stream-4 (b), stream-5 (c) and the total load (c) for a duration of five hours. The streams are detailed in Table \ref{fig:profile}.}
\vspace{-0.6cm}
\label{fig:Type_1_type_2}
\end{figure}

\textbf{Experiment under setting-2}: Second setting corresponds to a comparatively less loaded environment having 10 ACs and 10 water-heaters. The requests are also assumed to be arriving at a lesser rate. Here we also consider a set of 30 Type-1 devices from which requests come at a random fashion. Details are given in Table \ref{fig:profile} (\textit{stream-3}, \textit{stream-4} and \textit{stream-5}). Figure \ref{fig:Type_1_type_2}(a) and \ref{fig:Type_1_type_2}(b) depict the status of the system-load over a duration of 5 hr with and without application of the proposed load-management strategy. Figure \ref{fig:Type_1_type_2}(c) shows the load due to Type-1 devices. Figure \ref{fig:Type_1_type_2}(d) shows the combined system-load when all the streams (stream-3, stream-4 and stream-5) act together. In summary the proposed strategy successfully reduces the peak load upto 64\%  and the overall variation in the load upto 58\% throughout the experiment.

\begin{figure}[htbp]
\begin{center}
\includegraphics[angle=0,width=0.48\textwidth]{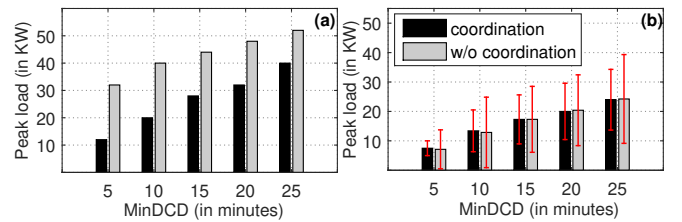}
\end{center}
\vspace{-0.3cm}
\caption{(a) Peak load and (b) average load (with standard deviation as the error bars) for different minDCD}
\label{fig:varing_minDCD}
\vspace{-0.3cm}
\end{figure}

\textbf{minDCD and maxDCP}: The duty-cycle characteristics of the Type-2 devices play a crucial role in the load-management algorithm. For a given maxDCP, the lesser the value of the minDCD, the more the number of the devices that can be accommodated together within a single period. Thus a lower value of minDCD provides better scope to flatten the system-load, and hence help to reduce the peak load too. On the contrary, as the minDCD rises, it becomes harder to accommodate multiple appliances in the same period which makes the peak load management difficult. We study this behaviour explicitly under a wide variety of experimental settings 
where the maxDCP is set to 30 minutes. Different experiments are done with different types of appliances having different minDCD starting from 5 mins to 25 mins. Figure \ref{fig:varing_minDCD}(a) and and Figure \ref{fig:varing_minDCD}(b) show how the peak load and the average load varies under different settings. A substantial rise in the average as well as peak load are visible with the increase in the minDCD. However, we find that proposed load-management strategy is able to reduce the peak load upto 62.5\%. Note that although the average load remains the same in both the cases, the abrupt variation in the load is largely reduced in all the cases as is evident through lesser standard deviation (error-bar) in Figure \ref{fig:varing_minDCD}(b). In particular, variation in the load subsides in the proposed strategy upto 62\%.

\begin{figure}[htbp]
\begin{center}
\includegraphics[angle=0,width=0.48\textwidth]{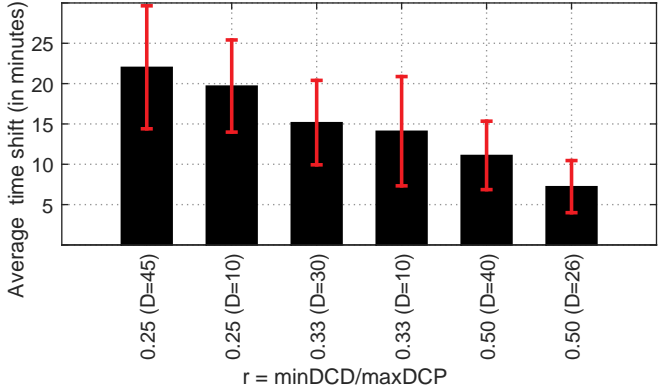}
\end{center}
\vspace{-0.3cm}
\caption{Average delay faced by the Type-2 devices in the proposed load-management strategy.}
\vspace{-0.1cm}
\label{fig:shift_profile}
\vspace{-0.3cm}
\end{figure}

\textbf{Time-delay}: We carry out experiments to understand the average delay that a Type-2 device would face, to get actually started from the time-point when the request was lodged. Such deviation would actually depend on the characteristics of the Type-2 devices, especially the maxDCP and the minDCD values. In particular, the larger the ratio of the minDCD and maxDCP, the lesser the number of devices that can be accommodated together in a single duty-cycle period. This in turn causes more devices to start together and hence less average waiting time. Similarly, the lesser the ratio, the higher the chance that a device need to wait longer to get started. Figure \ref{fig:shift_profile} shows the average time-delay for different values of the fraction $r =\frac{minDCD}{maxDCP}$. Average time-delay is lesser for higher $r$. However, delay goes higher for rise in the number of contenders also (D in X-axis of Figure \ref{fig:shift_profile}) as can be observed from the results.

\begin{figure}[htbp]
\begin{center}
\includegraphics[angle=0,width=0.48\textwidth]{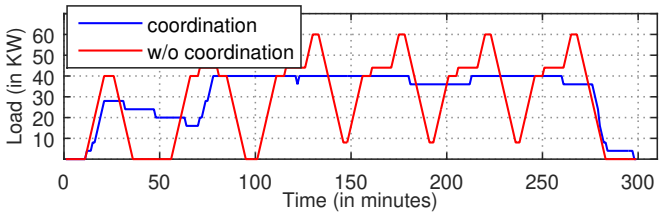}
\end{center}\vspace{-0.3cm}
\caption{Result for variation of system-load over a period of 5 hrs in DCube testbed.}
\vspace{-0.4cm}
\label{fig:dcube_results}
\end{figure}

\textbf{Evaluation over Testbed}: We rigorously tested the proposed load-management strategy in a publicly available IoT/WSN testbeds, DCube. We select a very dense cluster of the network having 30 IoT-nodes where all nodes are reachable to each other within one-hop. Each IoT-node is assumed to be representing the DI of attached to a Type-2 device. The requests arrive at a rate of 10 request/hour. The maxDCP and minDCD are assumed to be 30 and 15 min, respectively. Figure \ref{fig:dcube_results} shows the results with and without the proposed load-management strategy. In summary, the proposed strategy is found to reduce the peak load upto 33\% where the overall load variation subsides by 35\% throughout the experiment.

\vspace{-0.1cm}
\section{Conclusion and Future Works}
\label{sec:conclusion}

Efficient design and implementation of a HAN is a very fundamental issue in the context of a smart-grid. Most of the works done so far consider HAN as a centralized system. In order to get rid of the well-known problem with a centralized-solution, in this work we propose an IoT-based decentralized design of HAN. We exploit the advanced Synchronous-Transmission based IoT-protocols as the foundation of the proposed framework. We also propose a decentralized load-management strategy where the appliances can collaboratively decide a systematic execution plan/schedule in real-time. Through extensive experimental study in emulation and testbed settings we demonstrate that the proposed strategy can substantially bring down the peak system load as well as stabilize the overall system-load as much as possible. 

The cost of the reduction in the peak-load and overall smooth load curve achieved by the proposed load-management strategy is paid through a marginal delay the devices face in actually starting their execution. This specific issue can be easily resolved when a HAN is integrated with a smart-home system which can place the users-requests little ahead of time as per the user-profile. We plan to develop such sophisticated scheduling strategies as part of our future works in this direction.

\vspace{-0.2cm}

\bibliographystyle{unsrt}
\bibliography{smarthan}

\begin{thebibliography}{10}

\bibitem{ghosal2019key}
Amrita Ghosal and Mauro Conti.
\newblock Key management systems for smart grid advanced metering
  infrastructure: A survey.
\newblock {\em IEEE COMST, 2019}.

\bibitem{yan2012survey}
Ye~Yan, Yi~Qian, Hamid Sharif, and David Tipper.
\newblock A survey on smart grid communication infrastructures: Motivations,
  requirements and challenges.
\newblock {\em IEEE communications surveys \& tutorials, 2012}.

\bibitem{namboodiri2013toward}
Vinod Namboodiri, Visvakumar Aravinthan, Surya~Narayan Mohapatra, Babak Karimi,
  and Ward Jewell.
\newblock Toward a secure wireless-based home area network for metering in
  smart grids.
\newblock {\em IEEE Systems Journal, 2013}.

\bibitem{caprino2014peak}
Davide Caprino, Marco~L Della~Vedova, and Tullio Facchinetti.
\newblock Peak shaving through real-time scheduling of household appliances.
\newblock {\em Energy and Buildings, 2014}.

\bibitem{lee2011power}
Junghoon Lee, Gyung-Leen Park, Sang-Wook Kim, Hye-Jin Kim, and Chang~Oan Sung.
\newblock Power consumption scheduling for peak load reduction in smart grid
  homes.
\newblock In {\em ACM/SIGAPP SAC, 2011}.

\bibitem{6476768}
Xiaodao Chen, Tongquan Wei, and Shiyan Hu.
\newblock Uncertainty-aware household appliance scheduling considering dynamic
  electricity pricing in smart home.
\newblock {\em IEEE Transactions on Smart Grid, 2013}.

\bibitem{kim2011scheduling}
T{\`u}ng~T Kim and H~Vincent Poor.
\newblock Scheduling power consumption with price uncertainty.
\newblock {\em IEEE Transactions on Smart Grid, 2011}.

\bibitem{mohsenian2010optimal}
Amir-Hamed Mohsenian-Rad and Alberto Leon-Garcia.
\newblock Optimal residential load control with price prediction in real-time
  electricity pricing environments.
\newblock {\em IEEE transactions on Smart Grid, 2010}.

\bibitem{zimmerling2020synchronous}
Marco Zimmerling, Luca Mottola, and Silvia Santini.
\newblock Synchronous transmissions in low-power wireless: A survey of
  communication protocols and network services.
\newblock {\em ACM CSUR, 2020}.

\bibitem{missaoui2014managing}
Rim Missaoui, Hussein Joumaa, Stephane Ploix, and Seddik Bacha.
\newblock Managing energy smart homes according to energy prices: analysis of a
  building energy management system.
\newblock {\em Energy and Buildings, 2014}.

\bibitem{yoon2014dynamic}
Ji~Hoon Yoon, Ross Baldick, and Atila Novoselac.
\newblock Dynamic demand response controller based on real-time retail price
  for residential buildings.
\newblock {\em IEEE Transactions on Smart Grid, 2014}.

\bibitem{agnetis2013load}
Alessandro Agnetis, Gianluca De~Pascale, Paolo Detti, and Antonio Vicino.
\newblock Load scheduling for household energy consumption optimization.
\newblock {\em IEEE Transactions on Smart Grid, 2013}.

\bibitem{adika2013autonomous}
Christopher~O Adika and Lingfeng Wang.
\newblock Autonomous appliance scheduling for household energy management.
\newblock {\em IEEE transactions on smart grid, 2013}.

\bibitem{glossy}
Federico Ferrari, Marco Zimmerling, Lothar Thiele, and Olga Saukh.
\newblock Efficient network flooding and time synchronization with glossy.
\newblock In {\em ACM/IEEE IPSN, 2011}.

\bibitem{saha2017efficient}
Sudipta Saha, Olaf Landsiedel, and Mun~Choon Chan.
\newblock Efficient many-to-many data sharing using synchronous transmission
  and tdma.
\newblock In {\em IEEE DCOSS, 2017}.

\bibitem{leentvaar1976capture}
Krijn Leentvaar and Jan Flint.
\newblock The capture effect in fm receivers.
\newblock {\em IEEE Transactions on Communications, 1976}.

\bibitem{al2017network}
Beshr Al~Nahas, Simon Duquennoy, and Olaf Landsiedel.
\newblock Network bootstrapping and leader election utilizing the capture
  effect in low-power wireless networks.
\newblock In {\em ACM Sensys, 2017}.

\end{thebibliography}
\end{document}